\documentstyle[11pt,epsf,rotate,aaspp4]{article}

\received{}
\revised{}
\accepted{}
\cpright{}{}

\journalid{}{}
\articleid{}{}
\paperid{}

\cpright{}{}
\ccc{}

\slugcomment{To appear in ApJ}

\lefthead{Terquem, Papaloizou, Nelson \& Lin}
\righthead{On the tidal interaction in close binaries}

\begin{document}

\title{On the tidal interaction of a solar--type star with an orbiting
companion: Excitation of g~mode oscillation and orbital evolution}

\author{C. Terquem\altaffilmark{1,}\altaffilmark{3},
J.C.B. Papaloizou\altaffilmark{2}, R.P. Nelson\altaffilmark{2} and
D.N.C. Lin\altaffilmark{1}}

\altaffiltext{1}{UCO/Lick Observatory, University of California,
Santa--Cruz, CA~95064, USA -- ct,~lin@ucolick.org}

\altaffiltext{2}{Astronomy Unit, School of Mathematical Sciences,
Queen Mary~\& Westfield College, University of London, Mile End Road,
London E1~4NS, UK -- J.C.B.Papaloizou,~R.P.Nelson@qmw.ac.uk}

\altaffiltext{3}{{\it On leave from:} Laboratoire d'Astrophysique,
Observatoire de Grenoble, Universit\'e Joseph--Fourier/CNRS, BP~53,
38041 Grenoble Cedex 9, France}


\begin{abstract}

We calculate the dynamical tides raised on a non--rotating solar--type
star by a close stellar or planetary companion. Dissipation arising
from a turbulent viscosity operating in the convection zone and
radiative damping in the radiative core are considered.

We compute the torque exerted on the star by a companion in circular
orbit, and determine the potentially observable magnitude of the
tidally induced velocity at the stellar photosphere. These
calculations are compared with the results obtained by assuming that a
very small frequency limit can be taken in order to calculate the
tidal response ({\it equilibrium tide}). For a standard solar model,
the latter is found to give a relatively poor approximation at the
periods of interest of several days, even when the system is far from
resonance with a normal mode. This behavior is due to the small value
of the Brunt-V\"ais\"al\"a frequency in the interior regions of the
convection zone.

It is shown that although the companion may go through a succession of
resonances as it spirals in under the action of the tides, for a fixed
spectrum of normal modes its migration is controlled essentially by
the non--resonant interaction.

We find that the turbulent viscosity that is required to provide the
observed circularization rates of main sequence solar--type binaries
is about fifty times larger than that simply estimated from mixing
length theory for non--rotating stars. We discuss the means by which
this enhanced viscosity might be realized.

These calculations are applied to 51~Pegasi. We show that the
perturbed velocity induced by the tides at the stellar surface is too
small to be observed. This result is insensitive to the magnitude of
the turbulent viscosity assumed and is not affected by the possibility
of resonance. For this system, the stellar rotation and the orbital
motion are expected to be synchronized if the mass of the companion is
as much as one tenth of a solar mass.

\end{abstract}


\section{Introduction}
\label{sec:intro}

Theoretical analyses of the tidal interaction between close binaries
can be classified according to whether an equilibrium tide is assumed
or the dynamical tide is taken into account. The theory of the
equilibrium tide is based on the assumption that a star subject to the
tidal disturbance of a companion instantly adjusts to hydrostatic
equilibrium (Darwin~1879). A calculation including the dynamical tide
takes into account the fact that gravity or g~modes can be excited in
the convectively stable layers of the star and that resonances between
the tidal disturbance and the normal modes of the star can occur
(Cowling~1941). So far, dynamical tides have been studied only in
massive close binaries, which have a convective core and a radiative
envelope (Zahn~1975, 1977; Savonije~\& Papaloizou~1983, 1984, 1997;
Papaloizou~\& Savonije~1985, 1997; Savonije, Papaloizou~\&
Alberts~1995).

In this paper, we examine the effect of dynamical tides excited by a
companion on a solar--type star, in which a radiative core is
surrounded by a convective envelope.

This is of particular interest in connection with circularization of
solar--type binaries. It has been proposed that circularization occurs
through the action of turbulent viscosity, originating in the
convective envelope, on the tide. However, according to Claret~\&
Cunha~(1997) (see also Goodman~\& Oh~1997), who have used the
equilibrium tide formalism of Zahn~(1989), the circularization rate
resulting from this mechanism is too small by about two orders of
magnitude to account for the circularization timescales required on
the main sequence.

\noindent The tidal response calculation undertaken here is also of
interest in connection with the newly discovered planets, some of
which are found to orbit around solar--type stars with a period
comparable to that of the high order g~modes of the star. One such
example is 51~Pegasi (Mayor~\& Queloz~1995; Marcy~\& Butler~1995).

\noindent In these binaries, g~mode oscillations are excited by the
companion in the radiative region beneath the convective
envelope. They become evanescent in the convection zone where they are
damped by their interaction with the convective eddies. This
dissipation leads to an exchange of angular momentum between the star
and the orbit if the stellar rotation and the orbital motion are not
synchronized. Here we assume that the orbital frequency is initially
larger than the rotational frequency of the star.  Then tidal
interaction results in the decay of the orbit and the spin up of the
star. If the mass of the secondary companion is considerably smaller
than that of the primary, the timescale for orbital decay is smaller
than the stellar spin up timescale, and the companion eventually
plunges into the primary. But if the mass of the companion is large
enough, synchronization may occur before the binary has merged,
stopping further orbital decay. Estimates based on the theory of the
equilibrium tide (Rasio {\it et al.}~1996; Marcy {\it et al.}~1997)
suggest that the orbital decay timescale and the stellar spin up
timescale for a system like 51~Pegasi are longer than the inferred age
of the primary if the companion is a Jovian like planet.

In this paper we examine the effect of resonances on these timescales,
and determine the potentially observable magnitude of dynamical tides
at the photosphere of a solar--type star. We also compare the
dynamical tide calculations with the results of an asymptotic analysis
we carry out in the limit of small frequencies which should correspond
to the adiabatic equilibrium tide theory.

The paper is organized as follows: In \S\ref{sec:analysis}, we study
the tidal response of the star to the perturbation by a companion in a
circular orbit with a period in the range 4--13 days. In
\S~\ref{sec:linearized}, we first consider the linear adiabatic
response and then, away from resonance with a g~mode, extend the
analysis using first order perturbation theory to calculate the torque
due to dissipation in the convective envelope. This mechanism is then
found to be more important than non--adiabaticity arising from heat
transport in the radiative interior (i.e. radiative damping). However,
this is not the case in the vicinity of a g~mode resonance. There we
also calculate the torque due to non--adiabaticity in the radiative
core using a WKB treatment of the non--adiabatic terms. We find that
the torque at effective resonances is mainly determined by radiative
damping. An analysis valid for very low frequencies ({\it equilibrium
tide}) is given in \S~\ref{sec:asy}. 

\noindent The orbital circularization timescale for systems initially
in non--circular orbits can be derived from the response calculations
for companions in circular orbits. This is done in \S~\ref{sec:times}.
We then discuss how this might be used to calibrate the magnitude of
the turbulent viscosity required to fit the observations in
\S~\ref{sec:cal}.

\noindent Numerical calculations are presented in \S\ref{sec:results}.
The results assuming the equilibrium tide are given in
\S~\ref{sec:eqres}. In \S~\ref{sec:dyres} we present the results of
the dynamical tide calculations. We describe the tidal response of the
star to a companion in circular orbit, give the induced velocity at
its surface and the tidal torque. We describe the resonances and show
that, for the periods of interest of several days, they are not
expected to affect the orbital evolution of the binary. In
\S~\ref{sec:comp} we compare the calculations based on the dynamical
and equilibrium tides. We find that, for the standard solar model, at
the orbital periods of interest, because of the long timescale
associated with convection, the equilibrium tide calculations give a
relatively poor approximation to the results of the dynamical tide
calculations.

\noindent We find that the viscosity that is required to provide the
observed circularization rates is about 50 times larger than that
simply estimated from mixing length theory and discuss the means by
which this viscosity might be enhanced in \S~\ref{sec:calibration}.
However, we note that {\it the strength of the resonances for orbital
periods larger than $\sim 8$ days and the perturbed velocity at the
surface of the star are insensitive to the magnitude of the turbulent
viscosity assumed}. Only for periods $\sim 4$ days is the strength of
the resonances decreased by a factor $\sim 4$. The observable width of
the resonances as well is reduced when the viscosity is increased. We
also give the relation between the orbital evolution, circularization
and spin up timescales and the orbital frequency in
\S~\ref{sec:fitting}.

\noindent Finally in \S\ref{sec:discussion} we discuss and summarize
our results, applying them to 51~Pegasi in \S\ref{sec:peg}.

\section{Tidal response to a companion in circular orbit}
\label{sec:analysis}

The calculations presented in this section will be applied to close
binary systems where the primary is a solar--type star and the
secondary a stellar or planetary companion. The orbital periods of
interest lie in the range $4$--$13$ days. The rotational angular
velocity of the primary is assumed to be small compared to the orbital
frequency, so that it can be neglected. Quadrupolar tidal forcing thus
occurs through potential perturbations with periods in the range
$2$--$6.5$ days.

When calculating the tidal response well away from a condition of
resonance with a g~mode, we firstly calculate the tidal response
assuming it to be adiabatic throughout the star. First--order
perturbation theory is then used to calculate the dissipation
occurring in the convective envelope. The idea here (as is borne out
by the numerical results) is that although short wavelength g~modes
are excited in the radiative core, when they are away from resonance
they do not play an important role in comparison to the global
component of the tidal response. Also the variations in the convective
envelope occur on a comparatively long length scale, making the
adiabatic approximation a reasonable one.

When there is a resonance with a high order g~mode, the response
becomes one with a very short length scale such that non--adiabaticity
in the radiative core cannot be neglected. However, the modes are of
high order such that a WKB treatment of the non--adiabatic effects is
possible and this is used close to resonance where the normal mode
dominates the response. Such non--adiabatic effects turn out to be
more important than the action of turbulent viscosity in the
convective envelope, with the torque at significant resonances being
determined mainly by non--adiabatic effects.

\subsection{Linearized equations}
\label{sec:linearized}

\subsubsection{Adiabatic perturbations}

The linearized momentum, mass, and energy equations governing the
adiabatic response of the non--rotating star to the perturbing
potential $\Psi_T$ may be written (see, for example, Unno {\it et
al.}~1989)

\begin{eqnarray}
\frac{\partial^2 \mbox{\boldmath $\xi$}}{\partial t^2} & = & -
\frac{1}{\rho} \mbox{\boldmath $\nabla$} P' + \frac{\rho'}{\rho^2}
\mbox{\boldmath $\nabla$} P - \mbox{\boldmath $\nabla$} \Psi_T ,
\label{momentum} \\
\rho' & = & - \mbox{\boldmath $\nabla$} \cdot \left( \rho \mbox{\boldmath
$\xi$} \right) ,
\label{mass} \\
\Delta S & = & 0 ,
\label{energy}
\end{eqnarray}

\noindent where $P$ is the pressure, $\rho$ is the density, $S$ is the
entropy, $\mbox{\boldmath $\xi$}$ is the Lagrangian displacement
vector, $\Delta$ denotes the Lagrangian perturbation and the primed
quantities are Eulerian perturbations. We use the Cowling~(1941)
approximation, applicable to stars with high central condensation,
which neglects the perturbation to the stellar gravitational
potential. We also have the thermodynamic relation

\begin{equation} 
\Delta S = \frac{P}{\rho T} \frac{1}{\Gamma_3 -1} \left(
\frac{\Delta P}{P} - \Gamma_1 \frac{\Delta \rho}{\rho} \right) ,
\label{thermo}
\end{equation}

\noindent where $T$ is the temperature and $\Gamma_1$ and $\Gamma_3$
are the adiabatic exponents of Chandrasekhar. This relation together
with equation~(\ref{energy}) leads to

\begin{equation} 
\frac{\rho'}{\rho} = \frac{P'}{\Gamma_1 P} - A \xi_r,
\label{rhop}
\end{equation}

\noindent where 

\begin{equation}
A= \frac{dln\rho}{dr} - \frac{1}{\Gamma_1} \frac{dlnP}{dr} = -
\frac{N^2}{g} ,
\end{equation} 

\noindent with $sgn \left( N^2 \right) \times \sqrt{\left|N^2\right|}$
being the Brunt-V\"ais\"al\"a frequency and $g$ the acceleration  due to
gravity.

Following Cowling~(1941), only the dominant quadrupole term is
considered in the perturbing potential due to the companion. For a
binary system with a circular orbit, this is given in spherical polar
coordinates $(r,\theta,\varphi)$ by the real part of

\begin{equation}
\Psi_T \left( r, \theta, \varphi, t \right) = f r^2 \, Y_{n,m} \left(
\theta,\varphi \right) \,\, e^{- i m \omega t} ,
\label{eqpot} 
\end{equation}

\noindent where the spherical harmonic

\begin{displaymath}
Y_{n,m} \left( \theta,\varphi \right) = P^{|m|}_n (\cos\theta) \,\,
e^{i m \varphi}
\end{displaymath}

\noindent with $n=m=2$, $P^{|m|}_n$ being the associated Legendre
polynomial with indices $n$ and $m$. Here $\omega$ is the orbital
angular velocity, $f=-GM_p/4D^3,$ where $D$ is the orbital separation,
and $M_p$ is the mass of the companion. We adopt the same angular and
time dependence for the perturbations, so that $P',$ $\rho'$
and $S'$ are proportional to $Y_{n,m}(\theta,\varphi) \exp \left(- i m
\omega t \right)$. The corresponding expression for the Lagrangian
displacement is

\begin{equation} 
\mbox{\boldmath $\xi$} = \left[ \xi_r(r), \xi_h(r) {\partial \over
\partial \theta}, \xi_h(r) {\partial \over {\sin \theta \partial
\varphi} } \right] Y_{n,m}(\theta,\varphi) e^{- i m \omega t} .
\label{xi}
\end{equation} 

\noindent The factor $Y_{n,m}(\theta,\varphi) \exp \left(- i m \omega
t \right)$ will be henceforth taken as read, so that hereafter the
perturbations will be taken to depend only on $r$. Physical
perturbations are then found by taking real parts after inserting this
factor.

The horizontal displacement $\xi_h$ is given by the non--radial
equation of motion~(\ref{momentum}):

\begin{equation}
\xi_h = \frac{1}{m^2 \omega^2 r} \left( \frac{P'}{\rho} + f r^2
\right) .
\label{xih}
\end{equation} 

\noindent This relation together with equation~(\ref{rhop}) allow $P'$
and $\rho'$ to be eliminated from the
system~(\ref{momentum})--(\ref{energy}). The radial equation of
motion~(\ref{momentum}) and the continuity equation~(\ref{mass}) can
then be written as a  pair of ordinary differential equations for
$\xi_r$ and $\xi_h$:

\begin{eqnarray}
\frac{d \xi_r}{d r} & = & \left( - \frac{2}{r} + A - \frac{dln
\rho}{dr} \right) \xi_r + \left[ - \frac{m^2 \omega^2 r \rho}{\Gamma_1
P} + \frac{n(n+1)}{r} \right] \xi_h + \frac{f r^2 \rho}{\Gamma_1 P},
\label{ODE1} \\ \frac{d \xi_h}{dr} & = & \frac{1}{r} \left( 1 - \frac{AP}{m^2
\omega^2 \rho} \frac{dlnP}{dr} \right) \xi_r - \left( A + \frac{1}{r}
\right) \xi_h + \frac{Afr}{m^2 \omega^2} ,
\label{ODE2}
\end{eqnarray}

\noindent the solution of which requires two boundary conditions. At
the surface of the star we take a free boundary: $\Delta P=0$,
i.e. $P' =- \xi_r d P/dr.$ The boundary condition at $r=0$, where
equations~(\ref{ODE1}) and~(\ref{ODE2}) have a regular singularity, is
that the solutions be regular. Since at $r \sim 0,$ $P=P_c + O(r^2)$,
$\rho=\rho_c + O(r^2)$ and $A=O(r)$, where $P_c$ and $\rho_c$ are
respectively the central pressure and density, this leads to $\xi_h=
\xi_r /n$.

\subsubsection{Torque due to dissipation in the convective envelope}
\label{sec:torque}

The interaction between convective motions and the tidal flow is
expected to lead to the dissipation of tidally excited waves (e.g.
Zahn~1977). We model this effect as arising from a turbulent
viscosity. To do this we suppose there is an additional dissipative
force per unit mass acting in the convection zone given in spherical
coordinates by:

\begin{equation}
{\bf F}_c = \frac{1}{\rho r^2} \frac{\partial}{\partial r} \left( \rho
r^2 \nu \; \frac{\partial {\bf v}}{\partial r} \right) ,
\end{equation} 

\noindent where ${\bf v}$ is the flow velocity and $\nu$ is the
turbulent viscosity. Here, we assume that variation in the radial
direction is the most significant and note that the viscous force is
defined in such a way as to lead to a positive definite energy
dissipation rate.

\noindent For the turbulent viscosity coefficient, we take (see, for
example, Xiong, Cheng~\& Deng~1997)

\begin{equation}
\nu = \frac{c_1}{t_c} \frac{\Lambda^2}{1+c_2 \left( m t_c/P_o
\right)^s} ,
\label{nu}
\end{equation} 

\noindent where $c_1$ and $c_2$ are two constants, $c_1$ being on the
order of unity, $P_o=2\pi/\omega$ is the orbital period, $\Lambda$ is
the mixing length, and $t_c=1/\sqrt{|N^2|}$ is the convective
timescale. The viscosity is then $c_1 \Lambda^2/t_c$ for small forcing
frequency $m\omega$. The factor $1+ \left(m t_c/P_o \right)^s$, where
we shall use $s=2$, allows for a reduction of efficiency of the
damping of high frequency oscillations to which the convection cannot
adjust. A similar prescription with $s=1$ has been proposed by
Zahn~(1966) and used by Zahn~(1989), whereas $s=2$ has been considered
by Goldreich~\& Keeley~(1977) and used by Campbell~\&
Papaloizou~(1983) and Goldman~\& Mazeh~(1991). Goodman~\& Oh~(1997)
have also recently put forward some arguments in favor of $s=2$. For
the mixing length, we shall take the standard relation $\Lambda =
\alpha/ \left| dlnP/dr \right|$ and set $\alpha=3$.

In principle, equations~(\ref{momentum})--(\ref{energy}) should be
solved with ${\bf F}_c$ added on the right hand side of
equation~(\ref{momentum}) in the convective envelope. However, this
would increase the order of the differential system to be solved and
make the numerical calculations much more complicated. Instead, we
have found it adequate to solve first the adiabatic problem and then
to treat the dissipative effect using a first--order perturbation
theory. This is valid everywhere, except very close to resonances,
because dissipation is weak enough so that the imaginary parts of
$\xi_r$ and $\xi_h$ are much smaller in magnitude than their real
parts. Thus we solve equations~(\ref{momentum})--(\ref{energy})
without dissipative terms, and use these (real) solutions to calculate
the rate of energy dissipation $dE/dt$ due to convection which is
given by:

\begin{equation}
\frac{dE}{dt} = - \int_{V_c} \rho \; Re \left( {\bf F}_c \right) \cdot
Re \left( {\bf v} \right) \; dV,
\label{dEdt}
\end{equation} 

\noindent where the integration is over the volume $V_c$ of the
convective envelope and the angular dependence of ${\bf F}_c$ and
${\bf v}$ has to be taken into account. Using the relation ${\bf v} =
\partial \mbox{\boldmath $\xi$} / \partial t$ and the
expression~(\ref{xi}) for $\mbox{\boldmath $\xi$}$, we get

\begin{equation}
\frac{dE}{dt} = - \frac{48 \pi}{5} m^2 \omega^2 \int_{R_i}^{R_c}
\rho r^2 \nu \left[ \left( \frac{\partial \xi_r}{\partial r} \right)^2
+ 6 \left( \frac{\partial \xi_h}{\partial r} \right)^2 \right] dr ,
\label{PT1}
\end{equation} 

\noindent where $R_i$ and $R_c$ are respectively the inner and outer
radius of the convective envelope. Noting that the ratio of the rate
of exchange of energy and angular momentum with the orbit is given by
the pattern speed of the tidal disturbance, $\omega,$ the torque
exerted by the companion on the star is given by

\begin{equation}
{\cal T} = - \frac{1}{\omega} \frac{dE}{dt}.
\label{dET}
\end{equation} 

\noindent This torque is positive because the star is non--rotating.

When the frequency of the tidal wave is equal to that of some
adiabatic normal mode frequency of the star, we can no longer use
first--order perturbation theory because it would give an infinite
torque. However, when the frequency $\omega$ is very close to a
resonant frequency $\omega_0$, the torque will be given by an
expression of the form

\begin{equation}
{\cal T} = \frac{{\cal A}}{m^2 \left( \omega - \omega_0 \right)^2 +
\gamma^2},
\label{restor}
\end{equation} 

\noindent where ${\cal A}$ is an amplitude and $\gamma$ is the damping
rate for the mode. First--order perturbation theory assumes
dissipative effects are small in the response calculation. Therefore
it is valid only for frequencies such that $m^2 (\omega-\omega_0)^2
\gg \gamma^2$. However, the damping rate, if small, can be found from
first--order perturbation theory applied, as described above, very
close to resonance where the mode dominates the response. Then it is
given by (see, for example, Goldstein~1980)

\begin{equation}
2 \gamma = - \frac{1}{2 E_K} \frac{dE}{dt}
\label{gamma}
\end{equation} 

\noindent very close to resonance, where $E_K$ is the kinetic energy
of the mode:

\begin{equation}
E_K = \frac{1}{2} \int_V \rho \left[ Re(v) \right] ^2 dV = \frac{24
\pi}{5} m^2 \omega^2 \int_0^{R_{\sun}} \rho r^2 \left( \xi_r^2 + 6
\xi_h^2 \right) dr,
\end{equation} 

\noindent the integration being over the volume $V$ of the star (in
the first integral, the angular dependence of ${\bf v}$ has to be
taken into account). In~(\ref{gamma}), the total energy of the mode is
$2E_K$ because at resonance there is equipartition between kinetic and
potential energy. We calculate $\gamma$ using (\ref{gamma}) as
outlined above making sure that $\omega$ is close enough to $\omega_0$
by checking that $\gamma$ remains approximately constant when $\omega$
is slightly changed. To get ${\cal A},$ we fit the torque obtained
from first--order perturbation theory to the expression~(\ref{restor})
in the region approaching the resonance where still $m^2
(\omega-\omega_0)^2 \gg \gamma^2$. This procedure works satisfactorily
when $\gamma$ is small with consequent strong resonances. This appears
to be the situation when turbulent viscosity alone is assumed to
act. However, radiative damping cannot be neglected for resonances at
low forcing frequency and this is discussed below.

\subsubsection{Torque due to non--adiabaticity in the radiative core}
\label{sec:torqueNAD}

Non--adiabatic effects become important when the radiative diffusion
time across the length scale associated with the tidal response
shortens to become comparable to the wave propagation time across the
star. In principle, these effects should be taken into account both in
the radiative core and above the convection zone. However, since
g~modes, which are excited in the radiative core, are evanescent in
the convective envelope, we do not expect non--adiabaticity to play an
important role above the convection zone. To take into account
non--adiabatic effects, equation~(\ref{energy}) has to be modified in
the radiative core to:

\begin{equation}
\rho T \frac{\partial \left( \Delta S \right)}{\partial t} = -
\mbox{\boldmath $\nabla$} \cdot {\bf F}' ,
\label{NAD}
\end{equation}

\noindent where ${\bf F}'$ is the perturbed radiative
flux. The radiative flux is given by the radiative diffusion equation 

\begin{displaymath}
{\bf F} = - K \mbox{\boldmath $\nabla$} T,
\end{displaymath}

\noindent where $T$ is the temperature and $K=4acT^3/(3 \kappa \rho)$
is the radiative conductivity, with $a$ being the Stefan--Boltzmann
radiation constant, $c$ the velocity of light and $\kappa$ the opacity. 
Therefore 

\begin{equation}
\mbox{\boldmath $\nabla$} \cdot {\bf F}' = - \frac{1}{r^2}
\frac{\partial}{\partial r} \left[ r^2 \left(K \frac{\partial
T'}{\partial r} + K' \frac{dT}{dr} \right) \right] - \mbox{\boldmath
$\nabla$}_h^2 \left( KT' \right) ,
\label{grflux}
\end{equation}

\noindent where $\mbox{\boldmath $\nabla$}_h$ is the non--radial
component of the operator $\mbox{\boldmath $\nabla$}$. We now suppose
that close to resonance, the response behaves exactly like a free
g~mode with very large radial wavenumber $k_r$ so that WKB theory can
be used together with the local dispersion relation to evaluate
$\mbox{\boldmath $\nabla$} \cdot {\bf F}'$. The dominant term in the
right--hand side of~(\ref{grflux}) is then $-K \partial^2 T'/\partial
r^2=K k_r^2 T'$, and all the other terms can be neglected. For a high
order free g~mode, the local dispersion relation gives (see, for
example, Unno {\it et al.} 1989):

\begin{equation}
k_r^2= \frac{N^2}{(m \omega)^2} \frac{l(l+1)}{r^2} .
\end{equation}

\noindent This expression of $k_r$ is derived under the adiabatic
approximation. However, since we want to incorporate non--adiabatic
effects to the lowest order, we do not need to take them into account
in evaluating $k_r$. We have $k_r \gg 1/r$ in the radiative core
because $N \gg m \omega$ there. The perturbed temperature $T'$ can be
expressed as a function of $P'$ and $\rho'$ using the thermodynamic
relation

\begin{equation}
\frac{T'}{T} = \frac{1}{\chi_T} \frac{P'}{P} -
\frac{\chi_{\rho}}{\chi_T} \frac{\rho'}{\rho} ,
\end{equation}

\noindent where $\chi_T=\left. \partial lnP/\partial lnT
\right)_{\rho}$ and $\chi_{\rho}=\left. \partial lnP/\partial ln \rho
\right)_T$. 

\noindent The equation of state in the radiative core of a solar--type
star is primarly that of a perfect gas (we neglect here the radiation
pressure which is very small compared to the gas pressure). We then
have $\Gamma_1=\Gamma_3=5/3$ and $\chi_{\rho}=\chi_T=1$. Using the
above, the fact that $l=2$, $\partial/\partial t=-im \omega$ and the
relation~(\ref{thermo}), we can recast~(\ref{NAD}) under the form

\begin{equation}
\frac{\rho'}{\rho} = \frac{1+i \epsilon \Gamma_1}{1+i\epsilon}
\frac{P'}{\Gamma_1 P} - \frac{A}{1+i\epsilon} \xi_r ,
\label{NAD1}
\end{equation}

\noindent where 

\begin{equation}
\epsilon=\frac{16 a c T^4 N^2}{5 \left(m \omega \right)^3 \kappa \rho
P r^2} .
\end{equation}

\noindent For high order free g~modes, we have (see, for example, Unno
{\it et al.} 1989)

\begin{displaymath}
\frac{P'/ P }{A \xi_r} \sim \frac{m \omega}{N} \frac{dlnP}{dlnr} ,
\end{displaymath}

\noindent which means that $P'/P \ll A \xi_r$ in the radiative core of
a solar--type star at the frequencies of interest. The non--adiabatic
correction of the term associated with pressure in
equation~(\ref{NAD1}) is then very small compared to the
non--adiabatic correction of the term involving $\xi_r$. Therefore
equation~(\ref{NAD1}) can be approximated by

\begin{equation} 
\frac{\rho'}{\rho} = \frac{P'}{\Gamma_1 P} - {\bar A} \xi_r,
\label{rhopNAD}
\end{equation}

\noindent where we have defined ${\bar A}=A/(1+i\epsilon)$. We note
that because we have identified the tidal response with a normal mode,
this calculation is valid only close to
resonances. Equation~(\ref{rhopNAD}) is similar to~(\ref{rhop}) but
with $A$ being replaced by ${\bar A}$. The system of differential
equations we have to solve to get the non--adiabatic response of the
star is then the same as~(\ref{ODE1})--(\ref{ODE2}) but with the
following modification in the radiative core. In
equations~(\ref{ODE1}) and~(\ref{ODE2}), $A,$ where it appears as a
coefficient of $\xi_r$, has to be replaced by ${\bar A}$. The system
of differential equations so obtained is complex. In general, for the
periods we consider, $\epsilon \le 5 \times 10^{-4} $ is small. We
then calculate both the real and imaginary parts of the response, so
that the torque can be computed directly from:

\begin{equation}
{\cal T} = - \int_V \rho' \; {\partial \Psi_T\over \partial \varphi}
\, \, dV
\label{torgen}
\end{equation}

\noindent where the integral is over the volume $V$ of the star. The
angular dependence of $\Psi_T$ and $\rho'$ has to be taken into
account in this expression. Equation~(\ref{torgen}) can be recast
under the form (Savonije~\& Papaloizou~1983):

\begin{equation}
{\cal T} = - \frac{96 \pi f}{5} \int_0^{R_{\sun}} Im(\rho') r^4 dr .
\label{torqueNAD}
\end{equation}

As in \S~\ref{sec:torque}, we can calculate the damping rate $\gamma'$
in resonances due to non--adiabatic effects using~(\ref{gamma}).
Here, the rate of energy dissipation is calculated from the torque
(see~\ref{dET}).

\subsection{Low frequency limit}
\label{sec:asy}

In the limit of small $|\omega|$, the following relations may be
obtained for the {\it adiabatic equilibrium tide}:

\begin{equation}
P'_{eq} = - f r^2 \rho, 
\label{Peq}
\end{equation}

\noindent and, if $N^2 \ne 0$:

\begin{eqnarray}
\xi_{r,eq} & = & f r^2 \rho \left( \frac{dP}{dr} \right)^{-1} ,
\label{xieq} \\
\xi_{h,eq} & = & \frac{1}{n(n+1)r} \frac{d}{dr} (r^2 \xi_{r,eq} ),
\label{Weq}
\end{eqnarray}

\noindent where the `{\em eq}' subscript denotes the equilibrium
value. Using these displacements, the torque may be calculated using
equations~(\ref{PT1}) and~(\ref{dET}).

\noindent However, we comment that equation~(\ref{xieq}), which states
that the Lagrangian perturbation to the pressure is zero, can only be
derived in the adiabatic low $|\omega|$ limit if the
Brunt-V\"ais\"al\"a frequency is not zero (in practice, one also
requires that the forcing period be short compared to the thermal
timescale of the star, but the latter is so long that it can be
assumed to be infinite in this context).
Equations~(\ref{xieq})--(\ref{Weq}) do not apply in a finite region
where strictly $N^2=0.$ In that case the fluid is locally barotropic,
and the displacement can be written as the gradient of a potential:

\begin{equation}  
\mbox{\boldmath {$\xi$}}= \mbox{\boldmath $\nabla$} \left[ \Phi(r)
Y_{n,m} \right]. 
\end{equation}

\noindent The continuity equation then gives for low frequencies

\begin{equation}  
\mbox{\boldmath $\nabla$} \cdot \left( \rho \mbox{\boldmath {$\xi$}}
\right) = {1 \over r^2} {d \over dr} \left( r^2 \rho {d\Phi \over dr}
\right) - {n(n+1) \rho \Phi \over r^2} = - \rho'_{eq} = -{P'_{eq} \rho
\over \Gamma_1 P} = {f r^2 \rho^2\over \Gamma_1 P}.
\label{CRP} 
\end{equation}

\noindent Equation~(\ref{CRP}) gives a second order differential
equation for $\Phi(r).$ This applies inside the region where $N^2=0.$
It is possible, using the two available boundary conditions
for~(\ref{CRP}), to match $\xi_{r,eq}$ given by~(\ref{xieq}) at the
boundaries of such a region, but not in general $\xi_{h,eq}$ given
by~(\ref{Weq}). This means that there will tend to be a discontinuity
in the tangential displacement at the boundaries for low frequencies.

\noindent When $|N^2|$ is not zero but very small, in particular small
compared to $m^2\omega^2,$ which corresponds physically to the
convective timescale being much longer than the forcing period, the
tidal response more closely matches that given by~(\ref{CRP}) than
that given by equations~(\ref{xieq})--(\ref{Weq}). This feature causes
a very slow convergence towards the low frequency limiting solution
({\it equilibrium tide}) found here, as well as near discontinuous
behavior near the inner convective envelope boundary. This is borne
out by our numerical results (see \S~\ref{sec:velocity} and
\S~\ref{sec:comp}).

\subsection{Timescales}
\label{sec:times}

\subsubsection{Orbital evolution and stellar spin up timescales}

The torque, ${\cal T},$ gives the rate at which angular momentum is
transferred from the orbit to the star. We can then calculate a tidal
evolution (decay) timescale of the circular orbit:

\begin{equation}
t_{orb} = \frac{\mu \omega D^2}{{\cal T}},
\label{torb1}
\end{equation}

\noindent where $\mu=M_p M_{\sun}/(M_p + M_{\sun})$ is the reduced
mass. In principle, the variation of the torque with $\omega$ has to
be taken into account for the total decay time to be calculated. However,
since the torque increases as the companion spirals in, $t_{orb}$ is
mainly determined by the initial position of the companion, and a good
estimate can be obtained using the above formula.

This exchange of angular momentum also results in the spin up of the
star, the timescale of which is given by $t_{sp} = I \omega/{\cal T}$
(Savonije~\& Papaloizou~1983), with $I$ being the stellar moment of
inertia. Out of resonance, angular momentum deposition {\it initially}
occurs mainly in the convective envelope where the turbulent viscosity
acts (Goldreich~\& Nicholson~1989). It is then of interest to
calculate the spin up timescale for the convective envelope alone,
which is $t_{sp,c} = I_c \omega/{\cal T}$, where $I_c$ is the moment
of inertia of the convection zone. We note however that on the long
timescales associated with tidal evolution, angular momentum may be
redistributed throughout the star.

\subsubsection{Circularization}

In practice, we find that the torque arising from a companion in a
circular orbit varies with frequency approximately $\propto \omega^4.$
This result can be used to relate the orbital circularization
timescale to the initial orbital decay timescale provided that the
eccentricity is not too large. In practice, both these timescales can
be significantly longer than the spin up timescale of the star due to
its relatively small moment of inertia. We should then consider that
the star is synchronized with the orbit.

The ratio between the orbital decay timescale and the circularization
timescale $t_{circ}$ is found to be about 6 for the calculated
frequency dependence of the circular orbit torque (see, for example,
the expressions given in Savonije~\& Papaloizou~1983, 1984). This
appears to be independent of whether the star is assumed to be
synchronously rotating or non--rotating in that the circularization
rate calculated assuming no rotation, as we do here, gives a
reasonable estimate in the synchronous case also. In addition, to
evaluate $t_{circ}$ for an equal mass system, we have to take into
account the reciprocal torque exerted by the primary on the
companion. To do this, we take $t_{circ}$ to be proportional to a
factor which is $1/2$ when $M_p=M_{\sun}$ and 1 when $M_p \ll
M_{\sun}$. To a reasonable approximation, we then get:

\begin{equation}
t_{circ}=\frac{t_{orb}}{6 \left(1+M_p/M_{\sun} \right)} .
\label{tcircb}
\end{equation}

\subsection{Circularization timescale as a calibration of
turbulent viscosity}
\label{sec:cal}

One of the purposes of this paper is to calculate the tidally induced
velocities on the star. In order to do this, the processes responsible
for dissipating the disturbance should be included as accurately as
possible. An important dissipation process is that associated with
turbulent friction in the convection zone. As this process has been
suggested as being responsible for circularizing the orbits of main
sequence binaries of sufficiently short period (see Mathieu~1994), we
investigate whether reasonable assumptions about the behavior of
turbulent viscosity can lead to the observed circularization rates.

\subsubsection{Background}

Zahn~\& Bouchet~(1989) have investigated the pre--main sequence
evolution of late--type binaries in which the stars are fully
convective. The main conclusion of their work was that circularization
occurs at the very beginning of the Hayashi phase, with hardly any
decrease of the eccentricity on the main sequence. The cutoff period
they predict is between 7.3 and 8.5~days. According to them,
observations show a cutoff period around 8~days independent on the age
of the star and are then in agreement with their results.

However, we note that for this agreement to be reached, some
observations had to be discarded. Those by Mayor~\& Mermilliod~(1984),
which indicated that the cutoff period of late--type binaries was at
most 5.7~days, and those by Mathieu~\& Mazeh~(1988), which showed that
the cutoff period in the 4~$Gyr$ old cluster M~67 was more than
10~days. In a review article, Mathieu~(1994) (see also Mathieu~1992
and Mathieu {\it et al.}~1992) has confirmed that ``cutoff periods
increase with age, consistent with active main sequence tidal
circularization''. The pre--main sequence cutoff period is very likely
to be 4.3~$d$ (an upper limit being 6.4~$d$), and cutoff periods for
solar--mass binaries are 7.05~$d$ in the Pleiades (0.1~$Gyr$), 8.5~$d$
in the Hyades (0.8~$Gyr$), 12.4~$d$ in M~67 (4~$Gyr$) and 18.7~$d$ in
the halo field (16~$Gyr$). We therefore conclude that active
circularization does take place for main sequence binaries and that it
is less efficient than proposed by Zahn~\& Bouchet~(1989) on the
pre--main sequence. 

Recently, Claret~\& Cunha~(1997) have applied the formalism developed
by Zahn~(1989) to different stellar models. They have computed the
parameters which enter into the expression for the circularization
timescale, which is based on treatment of the equilibrium tide, for a
wide grid of stellar models as a function of mass and time. Their
conclusion is that turbulent dissipation is too low by a factor
100--200 during the main sequence to fit the observed cutoff periods.

\section{Numerical results}
\label{sec:results}

The calculations presented in the following section are applied to the
standard solar model described by Christensen--Dalsgaard {\it et
al.}~(1996).

\subsection{Equilibrium tide}
\label{sec:eqres}

The torque associated with the equilibrium tide was calculated as
indicated in \S~\ref{sec:asy}. As mentioned there, this calculation is
only expected to apply at very low frequencies. For periods between
4.23 and 12.4~$d$, and $c_1=c_2=1$, the calculated
 torque can be interpolated by
the following power law:

\begin{equation}
{\cal T} \; \left( g.cm^2/s^2 \right) = 1.200\times10^{55} \left(
\frac{M_p}{M_p+M_{\sun}} \right)^2 \omega^{4.08} .
\label{asyint}
\end{equation}

\subsection{Dynamical tide}
\label{sec:dyres}

For the calculation of the dynamical tide, we solve numerically the
differential equations~(\ref{ODE1}) and~(\ref{ODE2}) using a shooting
method to an intermediate fitting point (Press {\it et al.}~1986). To
evaluate non--adiabatic effects in the radiative core close to
resonances, we modify these equations in the way described in
\S~\ref{sec:torqueNAD}. We define the dimensionless quantity $x \equiv
r/R_c,$ where $R_c$ is the outer radius of the convective
envelope. With this notation, the equations are integrated from
$x_{in}=10^{-6}$ to $x_{out}=1.00071256.$ The radiative core extends
from $x=0$ to $x \simeq 0.7$. The results presented below
have been obtained with the values $c_1=c_2=1$ and $\alpha=3$
($\alpha$ being the ratio of the mixing length $\Lambda$ to the
pressure scale height). We discuss the effect of changing
these parameters.

\subsubsection{Tidal response and velocity at the surface of the star}
\label{sec:velocity}

\begin{figure}
\plotone{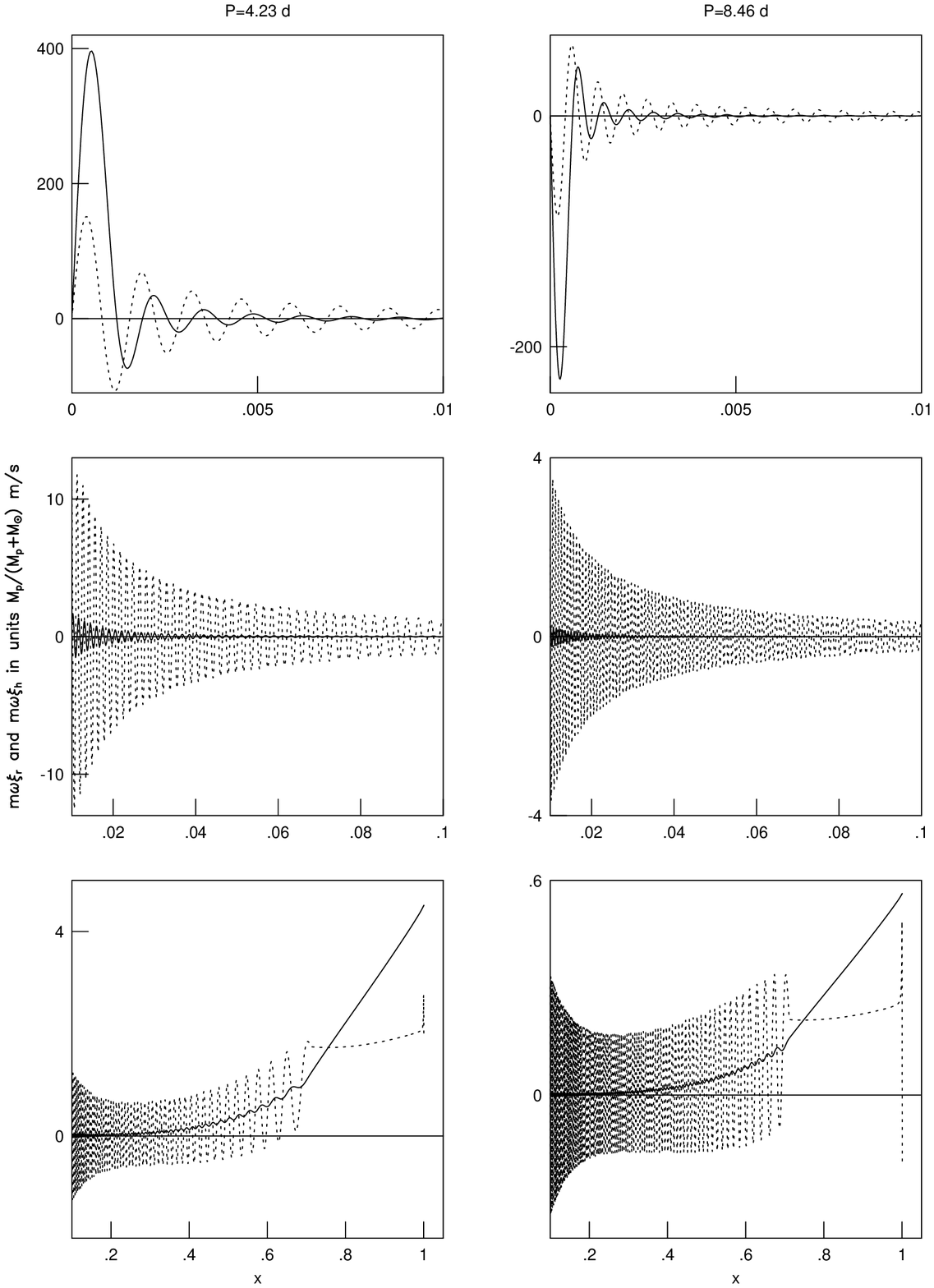}
\caption[]{Real part of $m \omega \xi_r$ (solid lines) and $m \omega
\xi_h$ (dotted lines) in units $M_p/(M_p+M_{\sun})$~$m/s$ versus $x$
for $x_{in} \le x \le 0.01$ (top panels), $0.01 \le x \le 0.1$ (middle
panels) and $0.1 \le x \le x_{out}$ (bottom panels), and for
$P_o=4.23$~$d$ (left panels) and $P_o=8.46$~$d$ (right panels). These
represent typical values of the radial and horizontal velocities, the
maximum values being three and six times larger respectively.}
\label{fig1}
\end{figure}

For illustration purposes, we plot the horizontal and radial
displacements induced in the star at orbital periods of $P_o=4.23$~$d$
and 8.46~$d$ away from resonance in the adiabatic approximation. The
first period is that inferred for the system 51~Pegasi. The spatial
distribution of the real parts of $m \omega \xi_r$ and $m \omega
\xi_h$ are shown in Figure~\ref{fig1}. These represent typical values
of the radial and horizontal velocities, the maximum values being
three and six times larger respectively. Since these quantities depend
on the perturbing mass through the ratio $M_p/(M_p+M_{\sun})$, they
have been represented in units of this factor. 

As expected, the stellar response shows oscillations between turning
points near the center and the inner radius of the convection zone,
where $m^2\omega^2=N^2,$ otherwise it is evanescent. The horizontal
displacement varies rapidly in the photosphere because the temperature
drops to zero rapidly there.

We see from Figure~\ref{fig1} that, when the perturbing mass
$M_p=M_{\sun}$, the maximum radial velocity at the surface of the star
is about 6 and 1~$m/s$ for $P_o=4.23$~$d$ and 8.46~$d$ respectively.
These numbers drop to $10^{-2}$ and $2\times10^{-3}$ respectively when
the perturbing mass is one Jupiter mass ($M_p=10^{-3}M_{\sun}$).

The radial displacement and the perturbed pressure at the surface of
the star are well approximated by the equilibrium values~(\ref{xieq})
and~(\ref{Peq}) respectively. These quantities are found to be
insensitive to the existence of resonances with the consequence that
the radial velocity at the surface of the star never differs much from
the values given above. For the smallest periods considered, the ratio
$\left| Re(\xi_h)/Re(\xi_r) \right|$ at the surface of the star can
vary by up to one order of magnitude on passage through
resonance. This is due to the fact that $\xi_h$ is proportional to
$\left(P'-P'_{eq}\right)/P$ (see~\ref{xih}). Even though this ratio
remains small, it can vary by up to an order of magnitude as a
resonance is passed through.

The numerical results indicate that both the amplitude and the
wavelength of the response increase with the orbital frequency, in
agreement with the theoretical expectation of a smaller radial order
for higher frequencies (Christensen--Dalsgaard~\& Berthomieu~1991 and
references therein).

Finally, as expected (see \S~\ref{sec:asy}), the plots shown in
Figure~\ref{fig1} (see also Figure~\ref{fig3}) indicate that at the
boundary of the radiative core and the convection zone there is a near
discontinuity in the mean value of $\xi_h$ obtained after averaging
out the interior oscillations.

\subsubsection{Circular orbit torque}
\label{sec:torquenum}

\begin{figure}
\plotone{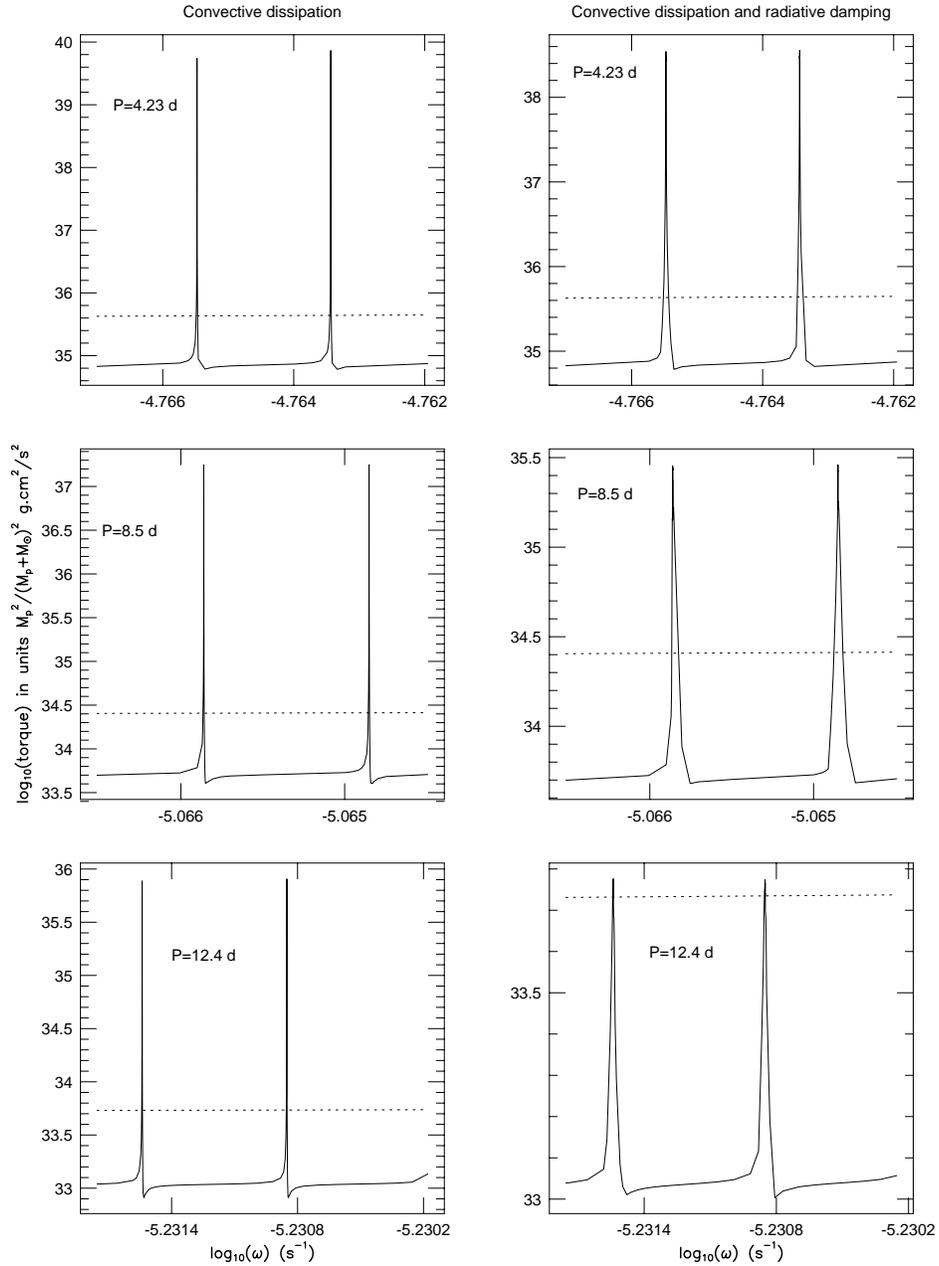}
\caption[]{$log_{10}({\cal T})$ with ${\cal T}$ in units
$M_p^2/(M_p+M_{\sun})^2$~$g.cm^2/s^2$ versus $log_{10}(\omega)$ for
$P_o=4.23$~$d$ (top panel), $8.5$~$d$ (middle panel) and $12.4$~$d$
(bottom panel). The solid and dotted lines correspond respectively to
the dynamical and equilibrium tides calculations. On the left panels,
${\cal T}$ is calculated using convective dissipation only. On the
right panels, ${\cal T}$ is calculated using radiative dissipation
alone in the resonances and convective dissipation alone away from
resonances.}
\label{fig2}
\end{figure}

\noindent {\it i) Resonances}:

Figure~\ref{fig2} shows ${\cal T}$ versus $\omega$ in a log--log
representation for three different small frequency intervals in the
vicinity of $P_o=4.23$, 8.5 and 12.4~$d$. On each plot, the dotted
line gives the values obtained from the theory of the equilibrium tide
as given by~(\ref{asyint}). Since the torque depends on the perturbing
mass through the factor $M_p^2/(M_p+M_{\sun})^2$, it has been
represented in units of this. These plots show several resonances,
which occur when the frequency of the tidal wave is equal to the
frequency of some normal mode of the star. The left panels show the
torque arising from convective dissipation, through turbulent
viscosity, alone (see \S~\ref{sec:torque}). These plots have been
displayed for comparison with the right panels, for which radiative
damping has been taken into account in the resonances (see
\S~\ref{sec:torqueNAD}). 

As indicated by the Table~\ref{table1}, the normal mode damping rate
due to radiative damping ($\gamma'$) is much larger than that due to
convective dissipation ($\gamma$).

\begin{table}[ht]
\footnotesize
\caption[]{Normal mode damping rates\\}
\begin{tabular}{ccccccccccccccccccccccccccccccccc}
\tableline
\tableline
$P_o$ & $\omega$ & $\gamma$ & $\gamma'$ \\
(days) & ($s^{-1}$) & ($s^{-1}$) & ($s^{-1}$) \\
\tableline
4.23 & $1.72\times10^{-5}$ & $6\times10^{-12}$ & $10^{-10}$\\
8.5  & $8.56\times10^{-6}$ & $5\times10^{-12}$ & $7\times10^{-10}$ \\
12.4 & $5.86\times10^{-6}$ & $4\times10^{-12}$ & $2\times10^{-9}$ \\
\tableline
\end{tabular}
\label{table1}
\end{table}

\noindent Thus the torque in the center of resonances, where they are
significant, is predominantly determined by radiative damping. For the
frequencies we consider, this contribution to the torque becomes much
smaller than that due to turbulent viscosity in the tail of the
resonances. This means that non--adiabatic effects in the radiative
core are negligible away from resonances. Therefore, in the right
panels of Figure~\ref{fig2}, we have plotted the torque resulting from
radiative damping acting alone in the center of resonances, and that
resulting from convective dissipation acting alone away from
resonances.  A comparison between the strength of the resonances shown
in the right and left panels indicates the importance of
non--adiabatic effects in the radiative core. As expected, the
resonances are weakened and broadened, this effect being marginally
important for $P_o \sim 4.23$~$d$.

We now discuss the properties and the effect of resonances on the
tidal torque. From now on, when resonances are discussed, we shall
refer to the calculations which take into account radiative damping.

\noindent In the neighborhood of $P_o=4.23$, 8.5 and 12.4~$d,$ the
relative separation, $\Delta\omega / \omega,$ between successive
resonances is respectively $4.5\times10^{-3},$ $2\times10^{-3}$ and
$10^{-3}.$ The relative width, $\delta\omega / \omega,$ of the
resonances is respectively $3\times10^{-4},$ $1.5\times10^{-4}$ and
$10^{-4}.$ Here we have arbitrarily defined the width of a resonance
as being the frequency interval over which ${\cal T}$ is at least 3
times larger than the minimum torque obtained just out of this
resonance.

\noindent To calculate the probability of the companion being in a
resonance, we have to take into account the fact that it drifts away
from the resonances much more rapidly than elsewhere. The relevant
quantity for calculating the tidal evolution timescale is $1/{\cal T}$
(see expression~\ref{torb1}). For a fixed oscillation spectrum, we can
approximate this probability by $\delta\omega/\Delta\omega$ times the
ratio of the mean value of $1/{\cal T}$ over a resonance to the mean
value of $1/{\cal T}$ between two resonances, where the mean value is
defined by

\begin{displaymath}
\left< \frac{1}{{\cal T}} \right> = \int \frac{d\omega}{{\cal T}}
\left/ \int d\omega \right. ,
\end{displaymath}

\noindent with the integrals being taken over the relevant frequency
interval.

\noindent This gives a probability of being in a resonance which is
close to 0.7\% for $P_o=4.23$ and 8.5~$d$ and 2\% for
$P_o=12.4$~$d$. The fact that the probability of being in a resonance
increases with $P_o$ is not significant, because resonances get weaker
when the period increases (see Figure~\ref{fig2}).

\noindent We note that this discussion applies only if the {\it a
priori} probability of being in any frequency interval of a given
width is independent of the frequency as might be expected to be a
reasonable assumption if the normal mode spectrum is fixed. However,
different circumstances may apply if the combined effect of orbital
and stellar evolution were to lock the companion in a resonance with
changing location. But we shall not consider the possibility of this
process here.

\noindent As the companion spirals inwards, it goes through a
succession of resonances. However, for a fixed normal mode spectrum,
the above calculation tells us that its migration is controlled
essentially by the non-resonant interaction. This can be seen by
comparing $<1/{\cal T}>$ evaluated over a large frequency range, both
taking into account and neglecting the resonances. Such a comparison
shows that neglecting resonances changes $<1/{\cal T}>$ by at most a
few per cent.

\noindent {\it ii) Relation between the mean torque, $\omega$ and the
circularization timescale}:

Here we interpolate the numerical results to express the torque as a
power of the frequency. To begin with, we consider the three frequency
intervals described above. We take the appropriate torque to be
$1/<1/{\cal T}>$ where the mean values are taken over the frequency
intervals displayed in Figure~\ref{fig2}. The results can be
interpolated with the following relation:

\begin{equation}
{\cal T} \; \left( g.cm^2/s^2 \right) = 1.654\times10^{53} \; \left(
\frac{M_p}{M_p+M_{\sun}} \right)^2 \; \omega^{3.85} .
\label{dynint}
\end{equation}

\noindent We have checked that the above formula gives a good estimate
of the torque at other non--resonant frequencies between 4.23 and
12.4~$d$.

\noindent Since the index of the power law~(\ref{dynint}) is close to
4, the circularization timescale $t_{circ}$ is given
by~(\ref{tcircb}). At $P_o=12.4$~$d$, $t_{circ}$ is found from the
above formula to be 56 times larger than the timescale of 4~$Gyr$ that
is indicated by the observations.

\noindent We note that both the dynamical and equilibrium tide
calculations give a power law with an index close to 4, which results
in the circularization timescale being proportional to the binary
period $P_o$ to the 13/3. For comparison, Zahn~(1977), Zahn~(1989) and
Goldman~\& Mazeh~(1991), using equilibrium tide calculations, found
$t_{circ}$ to be proportional to $P_o$ raised to the power 16/3, 13/3
and 10/3 respectively. The difference between these results can be
related to a different choice of $s$ in the expression~(\ref{nu}) for
$\nu.$ These authors used an expression similar to~(\ref{nu}) with
respectively $s=0$, 1 and 2. The fact that we obtain an index close
to 13/3 by setting $s=2$ or even $s=1$ (see below), in contrast to the
results above, is at least partially due to the effectively smaller
value of $c_2$ we used (see below).

\noindent We comment further that Tassoul~(1988) found $t_{circ}
\propto P_o^{49/12}$ for his postulated alternative hydrodynamical
mechanism for tidal circularization.
 
\subsection{Comparison between calculations based on the dynamical 
and equilibrium tides}
\label{sec:comp}
 
The results presented above show that the torque corresponding to the
dynamical tide is smaller than that given by the adiabatic equilibrium
tide for all the frequencies we have computed. However, the difference
tends to decrease as the frequency gets smaller. From
expressions~(\ref{asyint}) and~(\ref{dynint}), we calculate that the
ratio of these torques is indeed about 6.0 and 4.8 for $P_o=4.23$ and
12.4~$d$ respectively.
 
\begin{figure}
\plotone{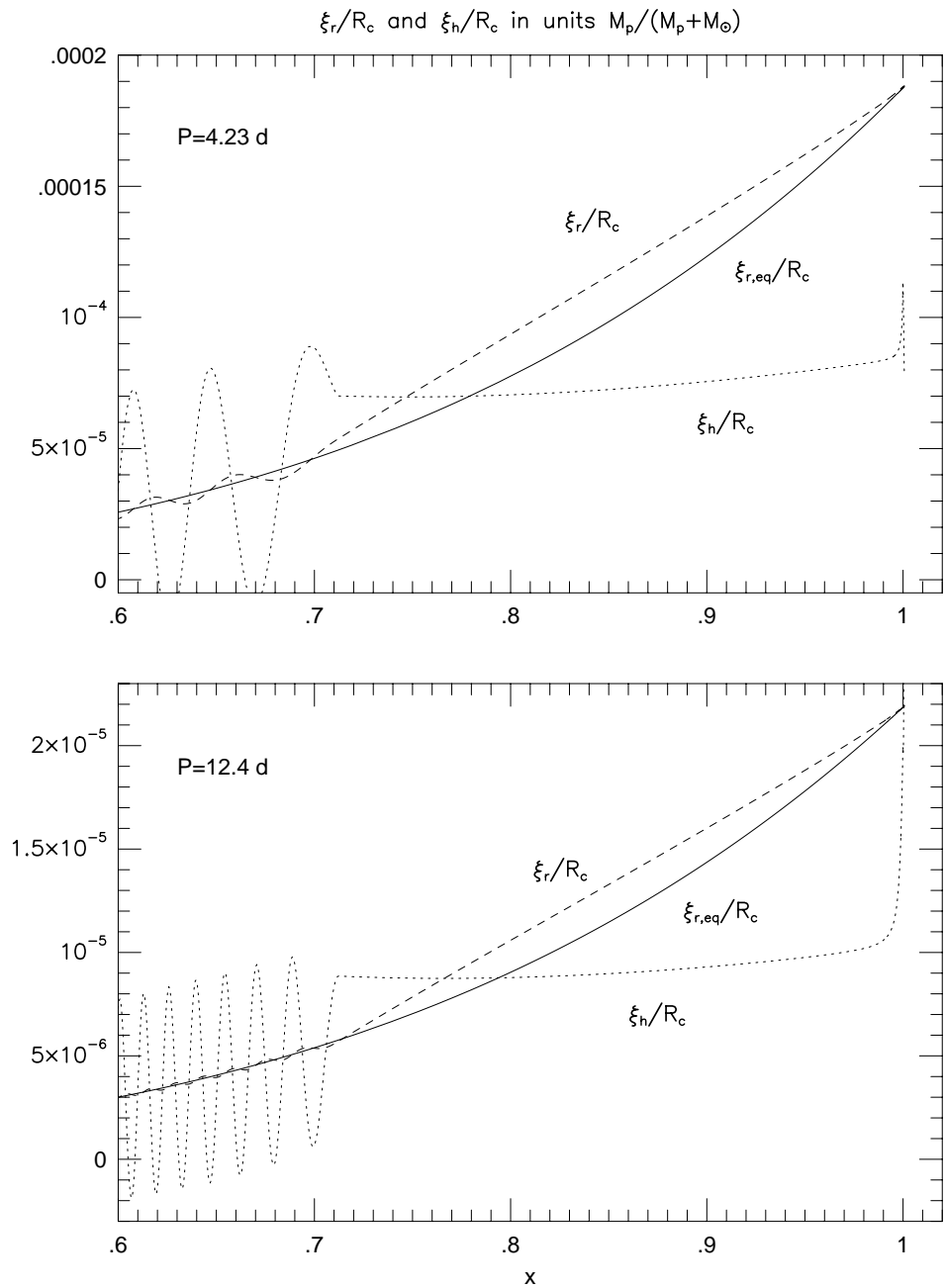}
\caption[]{$\xi_{r,eq}/R_c$ (solid line), $\xi_r/R_c$ (dashed line)
and $\xi_h/R_c$ (dotted line) in units $M_p/(M_p+M_{\sun})$ versus $x$
for $0.6 \le x \le x_{out}$, and for $P_o=4.23$~$d$ (top panel) and
12.4~$d$ (bottom panel).}
\label{fig3}
\end{figure}

\noindent Figure~\ref{fig3} shows $\xi_{r,eq}/R_c$, $\xi_r/R_c$ and
$\xi_h/R_c$ in units $M_p/(M_p+M_{\sun})$ versus $x$ in the range $0.6
\le x \le x_{out}$ for $P_o=4.23$ and 12.4~$d$. 

It is clear from these plots that $\xi_r$ departs from the asymptotic
value in the convective envelope. The difference is not large, being
about 17\%, but the derivatives of $\xi_r$ and $\xi_h$ from which the
torque is calculated (see~\ref{PT1}) depart more from their asymptotic
values.
 
\noindent In the limit where the magnitude of the Brunt-V\"ais\"al\"a
frequency is everywhere large compared to the tidal forcing frequency,
calculations based on the dynamical tide should converge towards those
based on the equilibrium tide. This is because the convective
timescale is small enough that the convective motions adjust
essentially instantaneously to the tidal forcing.

\begin{figure}
\plotone{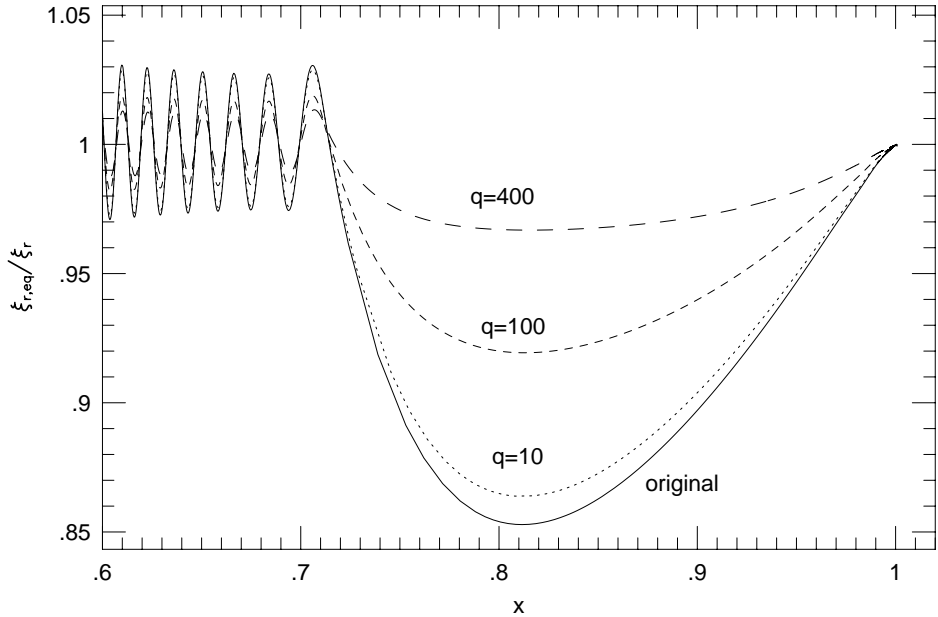}
\caption[]{$\xi_{r,eq}/\xi_r$ versus $x$ in the range $0.6
\le x \le x_{out}$ for $P_o=12.4$~$d$ and for $q=10$ (dotted line),
100 (short--dashed line) and 400 (long-dashed line), the definition of
which is given in the text. For comparison we have also plotted the
case corresponding to the original solar model (solid
line). $\xi_{r,eq}/\xi_r$ is close to 1 when $|N| \gg m\omega$.}
\label{fig4}
\end{figure}

\noindent We have checked this expectation by artificially increasing
the magnitude of the Brunt-V\"ais\"al\"a frequency in the convection
zone. Except in the part of the convective envelope just below its
outer radius, wherever $\left|N^2\right|<q \omega^2$, $q$ being an
arbitrary constant, we make the replacement $N^2 = -q \omega^2$. In
Figure~\ref{fig4} we plot $\xi_{r,eq}/\xi_r$ versus $x$ in the range
$0.6 \le x \le x_{out}$ for $P_o=12.4$~$d$ and for $q=10,$ 100 and
400. For comparison we also plot the case corresponding to the
original solar model. 

As expected, $\xi_{r,eq}/\xi_r$ converges towards 1 when the magnitude
of the Brunt-V\"ais\"al\"a frequency is increased. We note that
$q=400$, which corresponds to $|N| \ge 10 m \omega$, gives $1-
\xi_{r,eq}/\xi_r$ smaller than 5\%. This confirms that the asymptotic
limit is reached when the magnitude of the Brunt-V\"ais\"al\"a
frequency is very large compared to the frequency of the tidal wave.

We note that the very slow convergence towards the equilibrium tide
was predicted from the arguments presented in \S~\ref{sec:asy}. Also
expected was the discontinuity in the mean value of $\xi_h,$ obtained
after averaging out the interior oscillations, at the boundary of the
radiative core and the convection zone that is observed in
Figure~\ref{fig3} (see also Figure~\ref{fig1}).

\subsection{Calibration of the turbulent viscosity}
\label{sec:calibration}

We shall limit the comparison of our results with observations of main
sequence binaries because our calculations do not apply to pre--main
sequence stars, which have a much larger convective envelope than the
Sun. As mentioned above, the observed circularization timescale we
have to fit is then 4~$Gyr$ for $P_o=12.4$~$d$ (Mathieu~1994). As
indicated above, when using a simple estimate of the turbulent
viscosity based on mixing length theory for non--rotating stars, the
circularization timescale we get from our calculations for this period
is 56 times larger than 4~$Gyr$. This indicates that either \\ (i)
solar--type binaries are not circularized through turbulent viscosity
acting on tidal perturbations (but see Tassoul~1988 and Kumar~\&
Goodman~1996 for other suggested tidal mechanisms), or \\ (ii)
dissipation in the convective envelope of solar--like stars is
significantly more efficient than is currently estimated. 

Tassoul~(1995) postulates that efficient tidal dissipation occurs in a
very thin Eckman layer close to the surface of a tidally deformed star
and that this process greatly increases the efficiency of tidal
interactions. But a refutation of the notion that the free surface
boundary condition appropriate to the tidally deformed star, rather
than the more common rigid boundary condition, leads to such an
effective boundary layer, has been given by Rieutord~\&
Zahn~(1997). Further Tassoul~\& Tassoul~(1997) state that their
mechanism is inapplicable to extreme mass ratio cases such as
51~Pegasi that we consider later in the paper.

We now consider briefly here the mechanism proposed by Kumar~\&
Goodman~(1996), namely enhanced dissipation associated with high order
oscillation modes excited through parametric instability. The growth
rate for the most rapidly growing modes is expected to be $\sigma \sim
m \omega \xi_r/R_{\sun}$, where $\xi_r$ is the radial displacement in
the primary oscillation, which we shall assume to be the equilibrium
tide, evaluated at $r=R_{\sun}$, and $m \omega$ is its frequency.

\noindent If we assume that the non--linear development of the
parametric instability and subsequent dissipation of the excited modes
leads to an effective viscosity, and frictional dissipation rate
$t_f^{-1} =\sigma,$ which is big enough to suppress the linear
instability, then we expect from the classical theory of Darwin~(1879)
that there will be a phase lag $\theta_t$ associated with the tide
given by

\begin{equation}
\theta_t = \sigma \frac{R_{\sun}^3}{GM_{\sun}} \omega.
\label{darwin}
\end{equation}

\noindent For a binary of unit mass ratio and period $\sim 10$~$d$,
synchronization occurs on a timescale very much shorter than that
required for circularization, so that we assume that the stellar
rotation is synchronized with the orbit and $m=1$ in the calculation
of $\sigma.$ For small eccentricity, the circularization timescale
$t_{circ}$ is approximately given by $1/t_{circ} = \Omega_a \theta_t$,
where the apsidal motion frequency is $\Omega_a = 15 k (M_p/M_{\sun})
(R_{\sun}/D)^5 \omega$, with $k$ being the apsidal motion constant
(Cowling~1938). We use the equilibrium tide value~(\ref{xieq}) to
estimate $\xi_r/R_{\sun}$ at $r=R_{\sun}$ as $(R_{\sun}/D)^3 /4$ for a
mass ratio of unity. We then get $1/t_{circ}= 7.5 k \omega
(R_{\sun}/D)^{11} \propto \omega^{25/3}$ for $\theta_t$ given
by~(\ref{darwin}). For and orbital period of 12.4~$d$, this gives
$t_{circ}=6 \times 10^3/k$~$Gyr.$ Since $k\sim 0.01$, it is not very
likely that this mechanism will be able to explain the observed
circularization rates. However, we stress that this has not been shown
from a full non--linear calculation of the development of parametric
instability.

\subsubsection{Enhanced turbulent viscosity}

In general, a large increase in the simply estimated turbulent
viscosity coefficient is needed in order to explain the observed
circularization rate. We now investigate what is needed to achieve
this and give the numerical results of tests we have carried out. In
all cases, we have checked that {\it the velocity at the surface of
the star is not sensitive to the magnitude of the turbulent viscosity
assumed}.

In addition, as indicated above, the resonances are essentially
controlled by radiative damping in the radiative core as long as
$\gamma' \gg \gamma$. When the turbulent viscosity is enhanced,
$\gamma$ is increased. However, in the tests we present below, for
orbital periods larger than $\sim 8$ days, $\gamma'$ stays large
enough compared to $\gamma$ so that, although the non resonant torques
increase, the central structure and strength of the resonances is
determined by radiative damping. For orbital periods on the order of
$\sim 4$ days, $\gamma$ can become comparable to $\gamma'$. Then the
damping factor in~\ref{restor} has to be replaced by
$(\gamma+\gamma')^2$, and the strength of the resonance is reduced by
a factor $\sim 4$.

\noindent We first consider the effect on the circularization
timescale of {\it varying the parameters} $c_1$, $c_2$, $s$ and
$\Lambda$ in the expression~(\ref{nu}) for $\nu$.  Note that the
denominator we used in this expression is $1+c_2(mt_c/P_o)^2$ rather
than $1+c_2(m \omega t_c)^2.$ Using the latter with $c_2=1$ is
equivalent to setting $c_2 = (2\pi)^2$ in the former. At present, it
seems that our knowledge of convection does not allow discrimination
between these possibilities (see, for example, the discussion in
Zahn~1989). However, we note that Goodman~\& Oh~(1997) have recently
put forward some arguments in favor of $c_2 = (2\pi)^2$. Using $c_2 =
(2\pi)^2$ in~(\ref{nu}) results in a circularization time for
$P_o=12.4$~$d$ that is 10 times larger than that obtained with
$c_2=1$. If we set $c_2=0$, $t_{circ}$ is decreased by only a factor
2. This is because when $c_2=1$ the factor $c_2(mt_c/P_o)^2$ is
already smaller than, or even very small compared to, unity in a large
part of the convective envelope. Thus it seems that adjusting the way
in which turbulent viscosity responds to short period forcing cannot
produce the required enhancement in this case.

In Table~\ref{table2} we summarize the results obtained for different
values of $s$ and $c_2$, and in Table~\ref{table3} we indicate the
corresponding index of the power law in expression~(\ref{dynint}). We
note that $c_2=1$ corresponds to the denominator in~(\ref{nu}) being
$1+(mt_c/P_o)^s$, whereas $c_2=(2\pi)^s$ corresponds to $1+(m \omega
t_c)^s$.

\begin{table}[ht]
\footnotesize
\caption[]{$t_{circ}$ for different values of $s$ and $c_2$ \\}
\begin{tabular}{ccccccccccccccccccccccccccccccccc}
\tableline
\tableline
$P_o$ & $c_2$ & $s$ & $t_{circ}$ \\
(days) & & & ($Gyr$) \\
\tableline
4.23 & 1 & 2 & 2.46 \\
-- & -- & 1 & 1.37 \\
12.4 & -- & 2 & 220 \\
-- & -- & 1 & 211 \\
& & & \\
4.23 & $(2\pi)^2$ & 2 & 39.4 \\
-- & $2\pi$ & 1 & 5.98 \\
12.4 & $(2\pi)^2$ & 2 & 2010 \\
-- & $2\pi$ & 1 & 685 \\
\tableline
\end{tabular}
\label{table2}
\end{table}

\begin{table}[ht]
\footnotesize
\caption[]{Index of the power law in~(\ref{dynint}) for different
values of $s$ and $c_2$ \\}
\begin{tabular}{ccccccccccccccccccccccccccccccccc}
\tableline
\tableline
$c_2$ & $s$ & Index of the power law \\
\tableline
 1 & 2 & 3.85 \\
 -- & 1 & 4.3 \\
& & \\
$(2\pi)^2$ & 2 & 3.3 \\
$2\pi$ & 1 & 4.1 \\
\tableline
\end{tabular}
\label{table3}
\end{table}

\noindent We note that setting $c_2=(2\pi)^2$ with $s=2$ decreases the
index of the power law in~(\ref{dynint}) down to $\sim 3.3.$ This
gives $t_{circ}$ proportional to the orbital period to the $\sim
11/3$, which is similar to the value found by Goldman~\& Mazeh~(1991).

\noindent Zahn~\& Bouchet~(1989) have argued that the prescription
$s=2$ suggested by Goldreich~\& Keeley~(1977) and used later by
Campbell~\& Papaloizou~(1983) and Goldman~\& Mazeh~(1991) (see also
Goodman~\& Oh~1997) would lead to cutoff periods in clear conflict
with the observational data which they claim require $s=1.$ The
results presented for the model adopted here do not support this
statement. If $c_2=1$, $t_{circ}$ hardly changes when $s$ is changed
from 2 to 1. If $c_2=(2\pi)^s$, the difference between $s=1$ and $s=2$
is not dramatic for $P_o \sim 8-12$~$d$.  Within the uncertainties
associated with convection, our results do not allow a distinction to
be made between $s=1$ and $s=2$.

\noindent Finally, taking the mixing length to be the distance to the
top boundary of the convective envelope rather than 3 times the
pressure scale height does not affect significantly the
circularization timescale.

\noindent We note that the torque is directly proportional to $c_1$
and $\alpha^2$. However, $c_1$ is expected to be on the order of
unity, and $\alpha$ is usually taken to be between 1 and 4.

\subsubsection{Modifications to the Brunt-V\"ais\"al\"a
frequency}

The magnitude of the turbulent viscosity given by~(\ref{nu}) would be
increased if the magnitude of the Brunt-V\"ais\"al\"a frequency was
larger in the convection zone. This is because the convective
timescale decreases. We comment that a reduction in the convective
timescale, while maintaining the same length scale, implies larger
convective velocities which would have to occur without increasing the
heat flux. This is the essential feature of the modification. To
illustrate the effect of increasing the Brunt-V\"ais\"al\"a frequency
we consider the following dimensionless number:

\begin{displaymath}
\eta= \frac{N^2}{g} \left( \frac{dlnP}{dr} \right)^{-1},
\end{displaymath}

\noindent which is the superadiabatic temperature gradient when
radiation pressure and variations of the mean molecular weight are
neglected. The accuracy with which $\eta$ is known from helioseismic
observations is not better than $\sim 10^{-2}$ (Gough~1984). However,
in most of the convective envelope, this parameter, estimated from
mixing length theory applied to a non--rotating stars, is much smaller
than $10^{-2}.$

We have made a numerical investigation in which we increased
$\left|N^2\right|$ in the convection zone by replacing $\eta$ by min$(
p\eta ,10^{-3}),$ $p$ being an arbitrary constant, wherever
$\eta\le10^{-3}$ (except just below the outer radius of the convective
envelope). We have considered $p=50$ and $p=100$. It is doubtful that
consideration of present helioseismic data could preclude such an
increase of $\left|N^2\right|$ (Thompson~1997).

\begin{figure}
\plotone{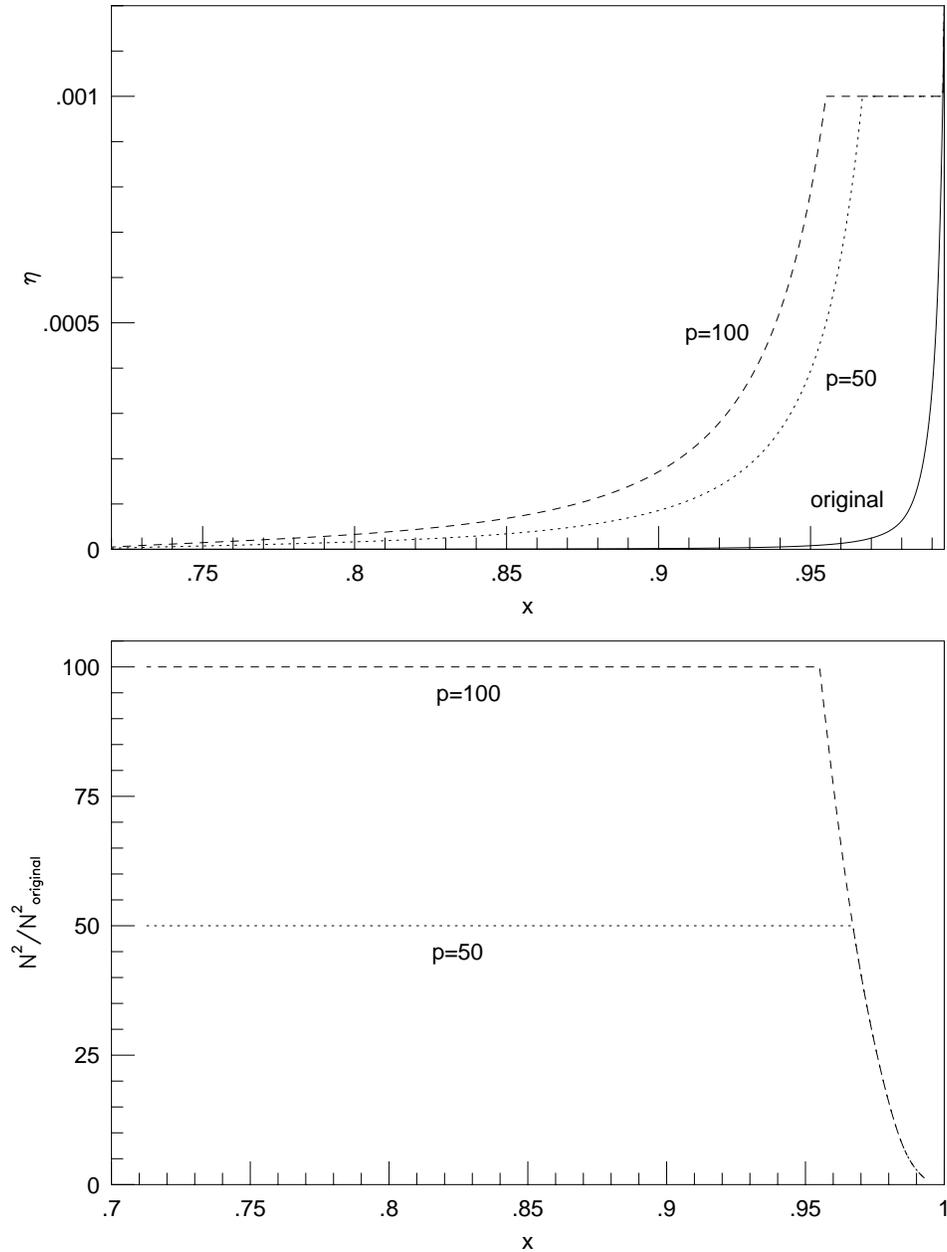}
\caption[]{{\it Top panel:} Dimensionless parameter
$\eta=N^2/(g \; dlnP/dr)$ in the convective envelope versus $x$. The
curves correspond to the original solar model (solid line), and to the
models with increased $\left|N^2\right|$ ($p=50$, dotted line, and
$p=100$, dashed line, where $p$ is defined in the text). {\it Bottom
panel:} Factor by wich $\left|N^2\right|$ is increased in the
convective envelope versus $x$. $N^2_{original}$ corresponds to the
original solar model, and $N^2$ corresponds to the models with $p=50$
(dotted line) and $p=100$ (dashed line).}
\label{fig5}
\end{figure}

\noindent Figure~\ref{fig5} shows $\eta$ in the convective envelope
versus $x$ for the original solar model and for $p=50$ and $p=100$. We
also display the factor by which $\left|N^2\right|$ has been increased
in each case.

\noindent The circularization timescales we find with $c_2=1$ and
$s=2$ for $P_o=12.4$~$d$ when $p=50$ and $p=100$ are respectively 13
and 7.7~$Gyr$, which are now larger than the observed value by factors
of 3 and 2 respectively. Small discrepancies of this magnitude could
be dealt with by adjustments to the mixing length or $c_1.$ When
either $p=50$ or $p=100$, the circular orbit torque is found to be
proportional to $\omega^{4.6}$.

If, keeping $s=2$, we adopt $c_2=(2\pi)^2$, $t_{circ}$ is increased
only by a factor 1.3 compared to the case $c_2=1$ for $p=100$. This is
because the factor $c_2(mt_c/P_o)^2$ is smaller than, or even very
small compared to, unity in almost all the convection zone whatever
$c_2$ between 1 and $(2\pi)^2$.

Although an increase in $\left| N^2 \right|$ of the magnitude we
consider might be thought to be unrealistic, we note that such an
increase in the deep layers of the convection zone has also been
considered by D'Silva~(1995) as a means of explaining the dynamics of
sunspots without invoking too strong a magnetic field. In his model,
which applies strictly to a star rotating at the same rate as the sun,
$\left| N^2 \right|$ has to be larger than $4 \times
10^{-11}$~$s^{-2}$, which means that it has to be multiplied on
average by a factor $\sim 8.$ In the solar model we use, we need to
multiply $\left| N^2 \right|$ by a factor between 100 and 400 for
$0.722 \ge x \ge 0.713$, between 10 and 100 for $0.828 \ge x \ge
0.722$ and between 1 and 10 for $0.915 \ge x \ge 0.828$ in order to
get such a minimum value. If we do this, the circularization timescale
we get for $P_o=12.4$~$d$ is only 6 times larger than the observed
one. In this context, it is possible that if the proposed increase in
$\left| N^2 \right|$ is related to the stellar rotation, this may be
even greater for the more rapidly synchronously rotating star that is
expected in the equal mass binary case.

We note that numerical simulations of turbulent convection in the
presence of rotation show an increase of $\left| N^2 \right|$ with the
effect of rotation (Brummell, Hurlburt~\& Toomre~1996). This is
because, as pointed out by Brummel {\it et al.}~(1996), rotation
influences the thermodynamic mixing properties of the convection in
such a way that it leads to a decrease in correlation between
temperature fluctuations and vertical velocities. The efficiency of
the vertical convective transport is then weakened, with a subsequent
enhanced superadiabatic mean stratification in the interior of the
fluid (see their Figure~8.a). \\ This suggests that the magnitude of
the Brunt-V\"ais\"al\"a frequency in the convective envelope of
rotating stars is actually larger than the values given by the solar
model we have been using here. However, it seems questionable that the
extremely large increase required to account for the observed
circularization rates can be achieved.

\subsubsection{Turbulent Viscosity below the convection zone}

Another means of increasing the total amount of dissipation is to
assume that turbulent viscosity acts down to some depth below the
inner boundary of the convective envelope. This might be expected if
convective overshooting takes place. However, this might not be a
very effective process because of the slow convective motions expected
and the rapid increase in $\left| N^2 \right|$ that occurs as the
radiative zone is entered.

Estimates based on the observed solar oscillation frequencies give an
upper limit between 0.05 (Basu~1997) and $\sim 0.1$
(Christensen--Dalsgaard, Monteiro~\& Thompson~1995) times the pressure
scale height on the extent of overshoot below the convection zone.

\noindent Here we consider a simple illustrative situation in which
convective blobs or some other turbulent motions are able to penetrate
into the stratified radiative core over some fraction $z$ of the
pressure scale height $\left| dlnP/dr \right|^{-1}$ producing a
turbulent viscosity. We model this by setting $\nu$ to be constant
from a distance $0.5 z \left| dlnP/dr \right|^{-1}$ above the inner
radius of the convective envelope down to the same distance below this
radius, equal to its value at the top of this zone. Because such a
viscosity is able to act on the short wavelength part of the tidal
response associated with g~modes it has a dramatic effect.

\noindent For $z=1$ and $c_1=c_2=1$, the circularization timescale for
$P_o=12.4$~$d$ is decreased by a factor about 400, being now 8 times
smaller than the observed timescale. If we set $c_2=(2\pi)^2$ (see
discussion above), $t_{circ}$ is increased by a factor 15, being about
4 times larger than the observed timescale.

\noindent For $z=0.1$ and $c_1=c_2=1$, we get a circularization
timescale for $P_o=12.4$~$d$ about 30 times larger than the observed
one. For the calculated timescale to be in agreement with the
observations, we need $z$ between 0.4 and 0.5 with $c_1=c_2=1$.  In
the model we have adopted, the effect is of less importance for
shorter periods because the number of oscillations of the response in
the region of the radiative core where turbulent dissipation is
introduced decreases with forcing frequency. Therefore, the torque
does not vary with $\omega$ as a simple power law with an index close
to 4. However, in reality, the effectiveness of the turbulent
viscosity should be reduced for short wavelength disturbances giving a
compensating effect to make it relatively less effective at low
frequencies.

The above calculations show that overshooting is not likely to be
efficient enough to decrease the circularization timescales by a
factor of about 50. To get such an effect, we indeed require an extent
of overshoot below the convection zone at least 5 or 10 times larger
than that deduced from the observations. In addition, we have not
taken into account the fact that overshooting leads to an increase of
the g~modes length scale through a decrease in the buoyancy or the
magnitude of $N^2.$ This in turn would decrease the amount of
turbulent dissipation associated with overshooting.

\subsection{ Fitting the observations}
\label{sec:fitting}

As we have already mentioned above, {\it calculations based on both
the dynamical and equilibrium tide theories give a torque proportional
to the orbital frequency raised to a power $\sim 4$} (see~\ref{asyint}
and~\ref{dynint}). If circularization of solar--type binaries does
occur through the action of turbulent viscosity on the tides, then its
magnitude has to be calibrated so as to account for the observed
timescale. We have discussed in the previous section, somewhat
speculatively, how the required enhancement of the magnitude of the
viscosity above that obtained from simple estimates might be envisaged
to occur.  Since the enhancement might depend on forcing frequency, it
is not clear that the resulting torque will still be proportional to
the frequency to the $\sim 4$. However, the increase to $\left| N^2
\right|$ described above gave torques that approximately preserved
this power law so that in the absence of additional information, we
shall suppose it holds. Then the calibration acts only to adjust the
coefficient of the power law.

\noindent We note that the observations do not rule out any exponent
between 3 and 5 (see below). Since our calculations can only strictly
be applied to solar--type stars, we calibrate our results using
$t_{circ}=4$~$Gyr$ for $P_o=12.4$~$d$. This gives (in cgs):

\begin{equation}
{\cal T} \; \left( g.cm^2/s^2 \right) = 5.086\times10^{35} \; \left(
\frac{M_p}{M_p+M_{\sun}} \right)^2 \; \left( \frac{\omega}{10^{-5} \;
s^{-1}} \right)^{4} ,
\label{T1}
\end{equation}

\noindent or, equivalently:

\begin{equation}
{\cal T} \; \left( g.cm^2/s^2 \right) = 1.423\times10^{39} \; \left(
\frac{M_p}{M_p+M_{\sun}} \right)^2 \; \left( \frac{P_o}{1 \; d}
\right)^{-4}.
\label{T2}
\end{equation}

\noindent The corresponding formul{\ae} for the orbital and spin up
timescales (in giga--years) are:

\begin{equation}
t_{orb} \; (Gyr) = 2.763\times10^{-4} \; \frac{\left( M_p/M_{\sun}+1
\right)^{5/3}}{M_p/M_{\sun}} \; \left( \frac{P_o}{1 \; d}
\right)^{13/3},
\label{torb}
\end{equation}

\begin{equation}
t_{sp} \; (Gyr) = 1.725\times10^{-6} \; \left(
\frac{M_p+M_{\sun}}{M_p} \right)^2 \; \left( \frac{P_o}{1 \; d}
\right)^{3} ,
\label{tsp}
\end{equation}

\noindent and $t_{sp,c}=I_c t_{sp}/I$ (for the solar model we use,
$I=1.064\times 10^{54}$~$g.cm^2$ and $I_c=1.5\times10^{53}$~$g.cm^2$).

\noindent Since the torque is proportional to $\omega^4,$
we can use the relation~(\ref{tcircb}) for $t_{circ}$, so that the
circularization time is given by

\begin{equation}
t_{circ} \; (Gyr) = 4.605\times10^{-5} \; \frac{\left( M_p/M_{\sun}+1
\right)^{2/3}}{M_p/M_{\sun}} \; \left( \frac{P_o}{1 \; d}
\right)^{13/3} .
\label{tcirc}
\end{equation}

\noindent Even though our calculations can only be applied to
solar--type stars, it is of interest to compare the circularization
timescales we get from~(\ref{tcirc}) with the observed ones. For
$P_o=4.3$, 7.05, 8.5 and 18.7~$d$, (\ref{tcirc}) gives respectively
$t_{circ}=0.04$, 0.3, 0.8 and 24~$Gyr$, to be compared with the
observed timescales 0.003, 0.1, 0.8 and 16~$Gyr$ respectively. The
agreement for $P_o \ge 8.5$~$d$ is within a factor 1.5. For smaller
periods, circularization is expected to occur when the convective
envelopes of the stars are larger, making turbulent dissipation more
efficient. We note that Mathieu {\it et al.}~(1992) have already
pointed out that a power law $t_{circ} \propto P_o^{13/3}$ provides a
close fit to the slope of the observed cutoff periods. However, the
observations are equally well fitted with an index of 10/3 and an
index of 16/3 cannot be ruled out (Mathieu {\it et al.}~1992).

\noindent If the simple estimate of the turbulent viscosity based on
mixing length theory for non--rotating stars is used, the coefficients
in the formul{\ae} for the torque have to be divided by a factor $\sim
50$ whereas those in the formul{\ae} for the timescales have to be
multiplied by the same factor.

\begin{figure}
\plotone{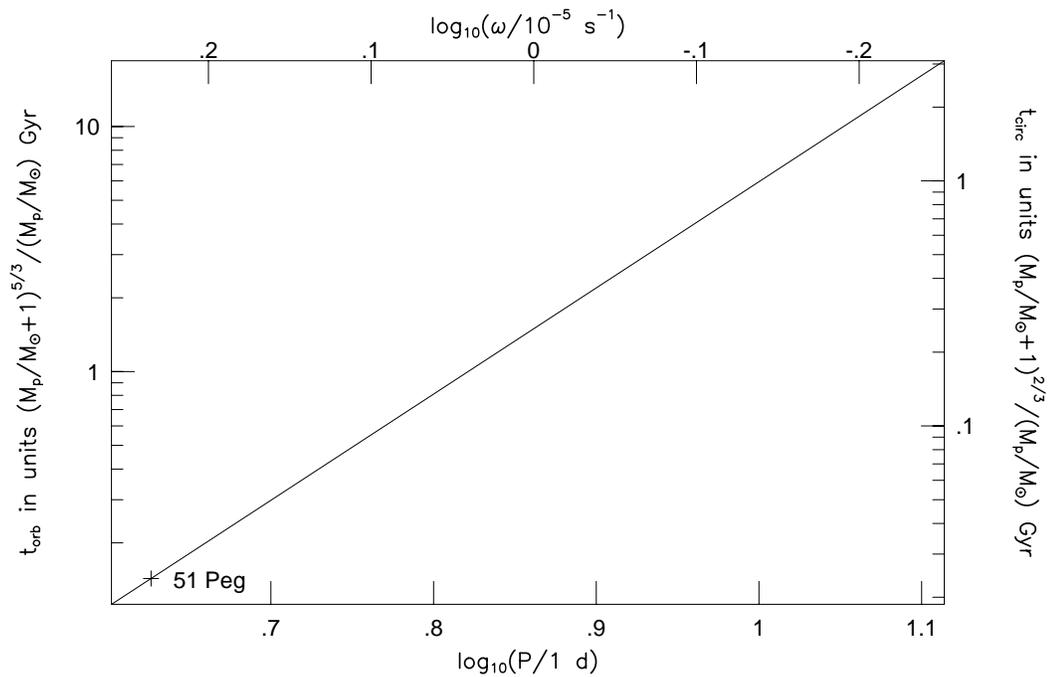}
\caption[]{$t_{orb}$ in units $\left( M_p/M_{\sun}+1
\right)^{5/3}/\left(M_p/M_{\sun}\right)$~$Gyr$ (expression~\ref{torb})
and $t_{circ}$ in units $\left( M_p/M_{\sun}+1
\right)^{2/3}/\left(M_p/M_{\sun}\right)$~$Gyr$
(expression~\ref{tcirc}) versus $\omega$ and $P$ in a log--log
representation. The cross indicates the position of 51~Pegasi. These
timescales fit the observations. {\it If instead they are calculated
using the simple estimate of the turbulent viscosity based on mixing
length theory, they have to be multiplied by $\sim 50$.}}
\label{fig6}
\end{figure}

\noindent In Figure~\ref{fig6}, we have plotted $t_{orb}$ in units
$\left( M_p/M_{\sun}+1 \right)^{5/3}/\left(M_p/M_{\sun}\right)$ and
$t_{circ}$ in units $\left( M_p/M_{\sun}+1
\right)^{2/3}/\left(M_p/M_{\sun}\right)$ versus $\omega$ and $P_o$ in
a log--log representation.

\section{Discussion }
\label{sec:discussion}

\subsection{Application to 51~Pegasi}
\label{sec:peg}

It is of interest to apply these results to the system 51~Pegasi, for
which the orbital period (assuming the observed oscillations are due
to a companion) is $P_o=4.23$~$d.$ If the companion is a Jupiter mass
planet ($M_p=10^{-3} M_{\sun}$), then the tidal orbital evolution
timescale given by~(\ref{torb}) is $t_{orb}\sim 140$~$Gyr,$ the star
spin up timescale~(\ref{tsp}) is $t_{sp}\sim 130$~$Gyr$ and the spin
up timescale of the convective envelope is $t_{sp,c}\sim
18$~$Gyr$. All of these timescales are long compared with the inferred
age of 51~Pegasi (Edvardsson {\it et al.}~1993). If the companion is a
low-mass star of $0.1 M_{\sun}$, as has been recently suggested,
$t_{orb}$ is 100 times smaller while $t_{sp}$ and $t_{sp,c}$ are
$10^4$ times smaller. We then expect the primary star to be
synchronized with the orbit, in which case exchange of angular
momentum is no longer taking place. Synchronization is actually
expected if the mass of the companion is larger than about 10 Jupiter
masses. The orbital decay timescale is also smaller than the age of
the system, but since $t_{sp} < t_{orb}$, tidal interaction stops
before the companion has plunged into the central star.

If the simple estimate of the turbulent viscosity based on mixing
length theory for non--rotating stars is used, all these timescales
have to multiplied by $\sim 50$. In that case, synchronization is
expected if the mass of the companion is larger than about 70 Jupiter
masses.

The planetary companion interpretation has been questioned recently by
the reported 4.23~$d$ modulation in the line profile of 51~Pegasi
(Gray~1997), and the possibility that this modulation may be due to
g~mode oscillations has been considered (Gray~\& Hatzes~1997).
 
We note that, according to our results, such a modulation could not be
due to g~mode oscillations tidally driven by a companion. For the
oscillation to have a period of $4.23$~$d$, the orbital period would
have to be 8.46~$d$. The maximum perturbed radial velocity at the
surface of the star induced by the companion would then be between
$2\times10^{-3}$ and 1~$m/s$ for a perturbing mass between $10^{-3}$
and 1 solar mass. {\it These numbers do not depend on the magnitude of
the turbulent viscosity assumed, and are not expected to be affected
by the possibility of resonance}. These velocities are at least about
50 times smaller than the observed ones.

\subsection{Summary}
\label{sec:sum}

In this paper, we have studied the dynamical response of the star to
the tidal perturbation of a companion. We have computed the torque due
to dissipation in the convective envelope using first--order
perturbation theory. In the vicinity of resonance, we have also
calculated the torque due to non--adiabaticity in the radiative core
using a WKB treatment. We have found that the torque at effective
resonances is mainly determined by radiative damping. We have carried
out an analysis based on the adiabatic equilibrium tide, and showed
that agreement with the dynamical tide calculations can be rather
poor. For the unmodified stellar model and the periods of interest of
several days, the torque derived using the equilibrium tide is 4 to 6
times larger than that corresponding to the dynamical tide.

We have found that the presence of fixed resonances do not affect the
long term orbital evolution of the binary, so that the different
timescales (orbital evolution, circularization and spin up) are mainly
determined by the non--resonant interaction. Our calculations show
that the viscosity that is required to provide the observed
circularization rates of solar--type binaries is $\sim 50$ times
larger than that simply estimated from mixing length theory for
non--rotating stars.

We have explored some means by which this viscosity might be
enhanced. We have found that it could become large enough if the
magnitude of the Brunt-V\"ais\"al\"a frequency in the deep convective
envelope were increased sufficiently. Such an increase is expected to
be produced by the effect of rotation on convection, but it is
questionable whether it can be of sufficient magnitude.

We note that {\it the strength of the resonances for orbital periods
larger than $\sim 8$ days and the perturbed velocity at the surface of
the star are insensitive to the magnitude of the turbulent viscosity
assumed}. Only for periods $\sim 4$ days is the strength of the
resonances decreased by a factor $\sim 4$. The effective widths of the
resonances affecting the tidal torques as well are reduced when the
viscosity is increased.

We have applied our results to 51~Pegasi, and showed that the
oscillations which have been observed at the surface of this star
cannot be a tidally driven non--radial g~mode. Also we have found that
the stellar rotation and the orbital motion of this system are
expected to be synchronized if the mass of the companion exceeds 0.1
solar mass.

\acknowledgments We are grateful to M.J. Thompson for supplying us a
solar model, and thank D.O. Gough, R.D. Mathieu and M.J. Thompson for
helpful discussions. This work is supported by PPARC through grant
GR/H/09454 and by NSF and NASA through grants AST~93--15578 and
NAG~5--4277. C.T. and D.N.C.L. acknowledge support by the Center for
Star Formation Studies at NASA/Ames Research Center and the University
of California at Berkeley and Santa-Cruz.

\newpage
\topmargin -1.cm

\end{document}